\documentstyle[epsfig,longtable]{aipproc}

\begin{document}

\title{Clustering of Ionizing Sources}
 
\author{Siang Peng Oh}
\address{Princeton University Observatory, Princeton, NJ 08544}

\maketitle

\begin{abstract}
Using existing models for the evolution of
clustering we show that the first sources are likely to be
highly biased and thus strongly clustered. We compute intensity
fluctuations for clustered sources in a uniform IGM. The strong
fluctuations imply the universe was reheated and reionised in a highly
inhomogeneous fashion.

\end{abstract}

\section*{Introduction}

Fluctuations in the ionizing intensity during reionization result in
temperature and ionization fraction fluctuations. This has important
implications for the reheating and reionization
history of the IGM. There are two contributing components:

${\bullet}$ IGM density fluctuations (variable optical depth)

${\bullet}$ Spatial distribution of ionizing sources

The second point is usually addressed by computing
Poisson fluctuations in the ionizing intensity (\cite{fardal},\cite{zuo}). Here we consider a
hitherto neglected effect, the clustering of ionizing sources, and
show that it is the dominant source of intensity fluctuations.

\section*{Modelling the Ionizing Sources}
We work within the framework of the concordance cosmology of
\cite{os95}:$(\Omega_{m},\Omega_{\Lambda},\Omega_{b},h,\sigma_{8
h^{-1}},n)= (0.35,0.65,0.04,0.65,0.87,0.96)$. We parametrise our model of reionization with 3 relevant lengthscales:

\medskip
\noindent
$\bullet${\bf Mean separation $n^{-1/3}$} The comoving number density of
collapsed objects above a critical mass ${\rm M_{*}}$ is given
by Press-Schechter theory: $n(z)= \Delta t \int_{M_{*}}^{\infty}
\dot{n}_{PS}(M,z) dM$, where the creation rate of dark matter halos is
given in \cite{sasaki}. The critical mass $M_{*}$ is given by the Jeans
mass after reheating, $M_{Jeans} \sim 10^{9}
(1+z/10)^{-3/2}M_{\odot} $. Before reheating, it is given by the virial mass necessary for efficient
atomic cooling $M_{cool} \sim 10^{8} (1+z/10)^{-3/2} M_{\odot} $ . The
source lifetime is assumed to be $\Delta t \sim 10^{7}$ yrs, the
characteristic timescale for both starbursts and quasars. 

\begin{figure}
\centerline{\epsfig{file=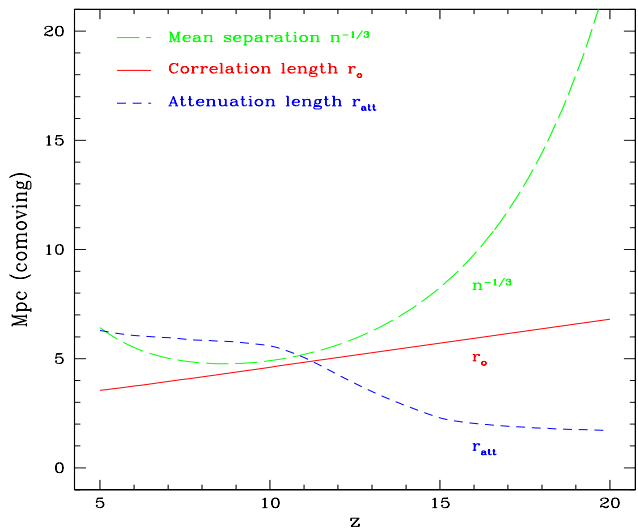,width=3.5in,height=3.0in}}
\vspace{10pt}
\caption{The evolution of various lengthscales with redshift. The increasing
bias factor with redshift counteracts the fall of the linear growth
factor, causing $r_{o}$ to remain almost constant. The downturn of
$r_{atten}$ at low redshift is caused by the increased clumping of the
gas, which shortens the recombination time.}\label{lengths}
\end{figure}
 
\medskip
\noindent
$\bullet$ {\bf Attenuation length $r_{\rm atten}$} This is modelled as
min($r_{strom},r_{life},r_{Lyman})$, where $r_{strom}$ is the Stromgren
radius, $r_{life}$ is the radius of the volume ionized by a source
during its lifetime, and $r_{Lyman}$ is the mean free path of photons
bounded by the covering factor of Lyman limit systems. We assume
values for the clumping of the IGM computed in \cite{go97}. The
covering factor of Lyman limit systems is computed by assuming a
mini-halo model (e.g.,\cite{abelmo}), and computing their abundance
via Press-Schechter theory.

\medskip
\noindent
$\bullet$ {\bf Correlation length $r_{o}$} The evolution of the matter
correlation function with redshift is given by integrating the power
spectrum. We assume a linear bias model $\xi_{\rm hh}(r)= b^{2}
\xi_{\rm mm}(r)$. Given the linear bias factor $b(M)$ as in \cite{mowhite}, we then compute $b(M,z)$ by assuming linear theory and
thus compute a number weighted bias factor at each epoch $\tilde{b}(z)$:
\begin{equation}
\tilde{b}(z)= \frac {\int_{M_{*}}^{\infty} b(M,z) n(M,z) dM} {\int
_{M_{*}}^{\infty} n(M,z) dM}
\end{equation}
The high bias means that the correlation length of halos is large,
even though the correlation length of matter falls with redshift. For
example, $\tilde{b} \sim 7$ at z=12, which leads to $r_{o}^{\rm halo}
\sim 3.3 {\rm Mpc} h^{-1}$ in comoving coordinates, only somewhat
weaker than the clustering of objects seen today. We
also calculate the evolution of the 3 and 4 point correlation
functions, assuming standard hierarchical scaling relations,
$\bar{\xi}_{N}(r) = {\rm S}_{N} \bar{\xi}_{2}^{N-1}(r)$. Note that the
$S_{N}$, although scale invariant, are a function of bias as well.

\begin{figure}
\centerline{\epsfig{file=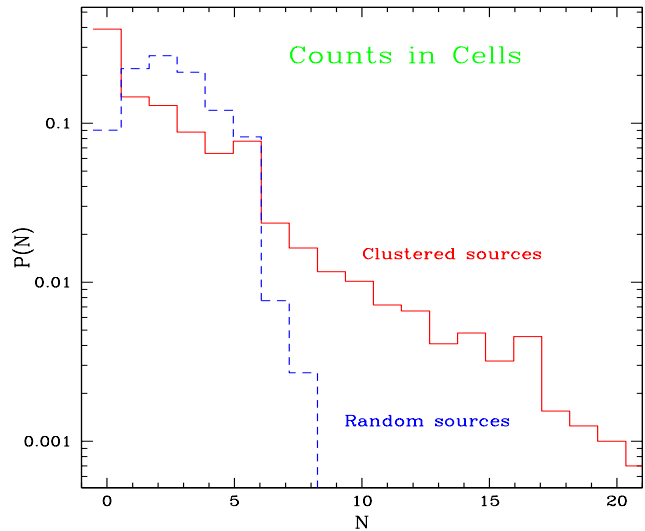,width=3.5in,height=3.0in}}
\vspace{10pt}
\caption{The probability distribution function of counts in cells at $z
\sim 12$ for clustered and Poisson distributed sources, for a sphere
of radius $r_{atten} \sim 4$ Mpc (comoving). The clustered case has a
high N tail, where many low luminosity sources can mimic an ultra
luminous source. Note also the greatly increased probability for N=0 (voids) in
the clustered case.}\label{countsprob}
\end{figure}

\medskip
\noindent
To estimate the fluctuations in the radiation field, we conduct a
Monte-Carlo simulation, in the spirit of \cite{fardal}. We generate a mock catalog of ionizing sources in a box 150 Mpc
comoving size at $z_{reion} \sim 12$. A fractal model \cite{soneira} is used to lay down sources with specified 2, 3 \& 4
point correlation functions, with appropriate modelling of the bias
factors. This has the chief advantage of speed and convenience over
using N-body simulations. It also does not suffer from mass
resolutions effects in identifying halos. We verify that the 2 point correlation function and the 2nd,
3rd and 4th moments of counts in cells (which are directly related to
the volume-averaged correlation functions, with corrections for shot
noise) are correctly recovered.
Luminosity is then randomly assigned to each source, assuming a Press-Schechter
mass function, and a constant emissivity per unit mass, as given by
\cite{haiman} for their mini-quasar model. The ionizing
flux at random points in the box is then computed, assuming each
source ionizes out to a radius $r_{\rm atten}$. We thus compute the
distribution function of the ionizing flux, as well as the two point
correlation function $\xi_{JJ}(r)=<(J(0)-<J>)(J(r)-<J>)>/<J>^{2}$.

\begin{figure} 
\centerline{\epsfig{file=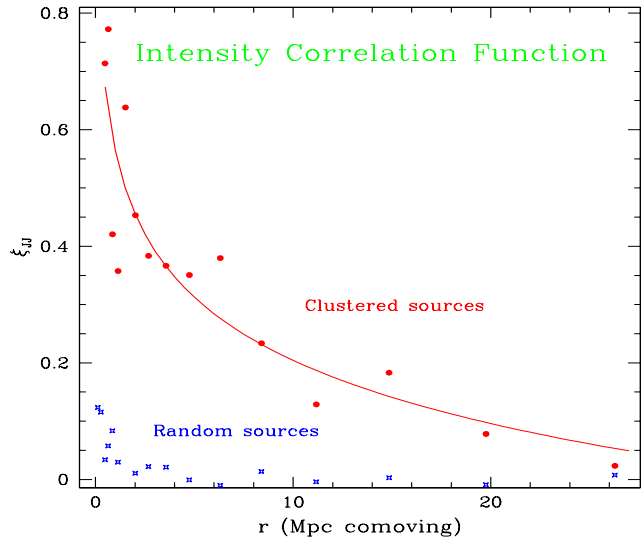,width=3.5in,height=3.in}}
\vspace{10pt}
\caption{The two-point correlation function of ionizing intensity at $z
\sim 12$ as a function of separation, assuming a sharp cutoff in
ionizing flux at $r_{atten}$. In the clustered case, the radiation
field displays correlations over large lengthscales. 
}\label{correl}
\end{figure}
\section*{Conclusions \& Observational Implications}

Clustering of ionizing sources produces a large dispersion in the
luminosity of random cells. Clustered sources collectively mimic an
ultra-luminous source. In addition, large voids exist where no
ionizing sources are present. These are ionized by the percolation of
ionizing radiation. Thus, large intensity fluctuations arise in the radiation field, which show coherence on large scales.

This has important observational implications. Emission line searches (e.g. in Ly$\alpha$ and H$\alpha$)
at high redshift have a stronger signal due to the higher intensity
fluctuations. Furthermore, CMB fluctuations due to Thomson scattering
off electrons in moving ionized patches \cite{knox} are enhanced, both due to the larger patch sizes and
the spatial correlations of the ionized patches. Finally, uneven
reheating provides a possible biasing mechanism.

{\bf Acknowledgements} I thank my advisor, David Spergel, for his
support and encouragement. I also thank Michael Strauss for helpful conversations.

\end{document}